# Design of Fuzzy self-tuning PID controller for pitch control system of aircraft autopilot

Nima Beygi[1,2*], Maani Beigy[2,3], Mehdi Siahi[1]

[1] Department of Electrical Engineering, Islamic Azad University, Garmsar Branch, Garmsar, IR Iran
[2] Power Electronic Laboratory, Department of Electrical Engineering, Tarbiat Modares University, Tehran, IR Iran
[3] Students' Scientific Research Center (SSRC), Tehran University of Medical Sciences, Tehran, Iran

## ABSTRACT

A variety of control systems have been proposed for aircraft autopilot systems. Traditional approaches such as proportional controller and conventional PID (CPID) controller are widely used. PID controller has a good static performance especially for linear and time-invariant systems, but a weak dynamic performance and discouraging function on nonlinear, time-varying, and uncertain systems. Fuzzy control theory can improve dynamic response in various conditions of system performance. This paper designs fuzzy self-tuning PID (FSPID) controller to improve disadvantages of conventional PID in aircraft autopilots. We apply proposed controller to pitch angle of aircraft then the abilities of proposed controller will be compared to the conventional PID and proportional controller. Inner feedback loop acts as oscillation damper in traditional schemes, but here is removed to compare the capabilities of Fuzzy self-tuning PID, conventional PID, and proportional controller. Based on the simulations, both of Conventional and Fuzzy self-tuning PID controllers can properly damp oscillations in lack of the inner feedback loop, but proportional controller cannot do. Then short-period approximation is assumed to assess the function of FSPID and CPID controllers in confront with abrupt and continuous disturbances, in addition to inappropriate tuning of parameters. Simulation results of short-period approximation show a better anti-disturbance function for Fuzzy self-tuning PID compare to the conventional type. Fuzzy self-tuning PID can tune the PID parameters for achieving the optimal response in view of speed, overshoot, and steady-state error in conditions of inappropriate tuning of PID parameters, based on the results of simulation in short-period approximation, the proposed controller can adaptively improve the system response by on-line setting of PID parameters.

**1. Introduction:**

Nowadays automatic control systems play a predominant role in civil and military aviation, so that various applications are used in modern aircrafts to help the flight crew in navigation, flight management, and augmenting the stability characteristics of aircraft (Barros dos Santos & de Oliveira, 2011; Wahid & Rahmat, 2010). Ordinary functions of manual aircraft guidance might be boring for pilots, which can be carried out by the autopilot systems. Autopilot assists the pilots in maintaining the route, heading or altitude, flying to navigation or landing references. In this way by just setting the target value that aircraft must be arrived in, autopilot controls it (Barros dos Santos & de Oliveira, 2011; Wahid & Rahmat, 2010). Various control approaches have been introduced for autopilots in the literature. Firstly non-linear control approach was used for flight control applications (Azam & Singh, 1994; Bugajski & Enns, 1992; Menon, Badget, Walker, & Duke, 1987; Tahk, Briggs, & Menon, 1986 ). Then Fuzzy logic control method was introduced in autopilot systems to improve nonlinear control imperfections (Bossert & Cohen, 2002; Cohen & Bossert, 2003; Kadmiry & Driankov, 2004; Wu, Engelen, Babuska, Chu, & Mulder, 2003).

Also linear models have been used in autopilots, which improve only parametric robustness, but have no excellence over other advantages of fuzzy non-linear approaches (Barkana, 2005; Cohen & Bossert, 2003). A useful overview about autopilot control systems has been done in the study of Babaei (Babaei, Mortazavi, & Moradi, 2011).

The first and the most applicable control strategy in the industry and engineering applications is PID control. Popularity of PID controller can be justified by some of its benefits: good performance, simple designing technique, robustness, and reliability (Haifang, Yu, & Tao, 2010; He, Jia, Li, & Gao, 2006; Hongbing, 2010). Also it has a good static performance especially for linear and time-invariant systems. Despite these benefits, conventional PID (CPID) controller has a weak dynamic performance. Its function on nonlinear, time-varying, and uncertain systems is not desirable. Industrial systems encounter disturbance and time-varying parameters, which result in imprecise function of CPID (Guo & Tang, 2009; Haifang, et al., 2010; Ming-shan, Yuan, Zi-da, & Li-peng, 2009; X.-k. Wang, Sun, Wanglei, & Feng, 2008). The CPID depends on precise mathematical model and transfer function of system, which is difficult to obtain in complex systems (Haifang, et al., 2010; Shoujun & Weiguo, 2006). The problem of adjusting PID parameters has been considered in previous studies, so that several methodologies are available in setting gains of PID controllers such as classical (Ziegler/Nichols, Cohen-Coon, pole placement and optimization, etc) or advanced techniques (minimum variance, gain scheduling and predictive). These methodologies have some disadvantages: 1) immoderate number of rules for adjusting the gains of PID, 2) improper function in nonlinear and uncertain systems with long time delay, and 3) mathematical complexity in tuning process (Chang, 2007; Iruthayarajan & Baskar, 2009; Ming-shan, et al., 2009; Sumar, Coelho, & Coelho, 2005).

---

*Corresponding author. Tel.: +98 21 8288 3976
Email addresses: nimabeigy@gmail.com (N Beygi),
m-beigy@student.tums.ac.ir (M Beigy),
mehdi_siahi@yahoo.com (M Siahi)

**Table 1 Nomenclature**

| Nomenclature | |
|---|---|
| $m$ | Mass |
| $g$ | Acceleration of gravity |
| $P, Q, R$ | Angular velocity components about x, y and z axes |
| $U, V, W$ | Components of velocity along the x, y and z axes |
| $L, M, N$ | Components of moment along the x, y and z axes |
| $I_x, I_y, I_z$ | Mass moments of inertia of the body about x, y and z axes |
| $I_{xy}, I_{yz}, I_{xz}$ | Products of inertia |
| $\theta, \Phi, \psi$ | Pitch, roll and yaw angles |
| $\delta$ | Control surface deflection angle |

The most important problem of CPID in the process of tuning is necessity to offline setting of parameters. Moreover PID controllers use fixed parameters in different conditions, which do not conduct optimal control (Ming-shan, et al., 2009; Truong & Ahn, 2011; Yuzhi & Haihua, 2008).

Fuzzy logic control has been reforming the inefficacy of PID controllers in non-linear, uncertain, and time-varying systems. Fuzzy control is an intelligent and non-linear control strategy, which employs fuzzy linguistic rules and does not need precise mathematical model of system. This strategy improves dynamic response considering a wide parameter variation in several conditions of system performance (Alp & Akyürek, 2011; Guo & Tang, 2009; Haifang, et al., 2010; Ming-shan, et al., 2009; Soyguder, Karakose, & Alli, 2009; X.-k. Wang, et al., 2008; Yongbin, Yongxin, & Cun, 2010). A variety of Fuzzy logic control approaches have been introduced in industrial applications, that can be categorized into: 1) Conventional Fuzzy control; 2) Fuzzy PID control; 3) neuro-Fuzzy Control; 4) Fuzzy-sliding mode control; 5) Adaptive Fuzzy Control; and 6) Takagi-Sugeno model-based Fuzzy Control (Feng, 2006). The conventional Fuzzy controller is a heuristic and model free method that was introduced by Mamdani and Assilian (Mamdani, 1974; Mamdani & Assilian, 1975). Fuzzy PID controller that was suggested by Bao-Gang et al (Bao-Gang, Mann, & Gosine, 2001), uses fuzzy control within conventional PID controller, and can also be classified as the *direct-action* type of fuzzy controller. By combining CPID control strategy and conventional Fuzzy control, a better control system can be achieved (Feng, 2006). Capabilities of neuro control in learning plus high computation efficacy of hybrid neuro-fuzzy strategies produce a powerful control system that is capable in data processing; which is more flexible, adaptive, and robust to the external disturbances or system variations (Feng, 2006). The integration of fuzzy control and modified sliding mode control results in an efficient controller. Although fuzzy control is an extension of sliding mode (Palm, 1992), the supervisory function of slide-mode in hybrid fuzzy-sliding mode provides stability and robustness of the closed loop control system (Feng, 2006). Adaptive control systems have the major problem of indispensable mathematical modeling especially in complex systems. Fuzzy control does not need precise mathematical modeling of system, thus it can overcome this problem of adaptive controllers (Feng, 2006; L. X. Wang, 1993). The dynamic model-based Fuzzy Control (Takagi & Sugeno, 1985) provides a basis for development of systematic approaches in stability analysis and controller design of fuzzy control systems, in terms of powerful conventional control theory (Feng, 2006).

This study aims to improve mentioned limitations of CPID by combining PID and Fuzzy controllers for aircraft autopilot. Fuzzy self-tuning PID (FSPID) is a hybrid controller identified with the ability to adaptive and online tuning of PID parameters in varying conditions of system function. In this paper we adopted FSPID to control the pitch angle of aircraft.

This paper is composed of five sections. In Section 2, flight principles and mathematical modeling of pitch control are explained. In Section 3, firstly the structure of FSPID is explained. Secondly the automatic parameter-tuning rules of PID controller are explained. Then membership functions and fuzzy rules are determined. In the final part of design section, our proposed controller is applied in the autopilot system. Section 4 shows the simulation results of our study. Concluding remarks are prepared in Section 5.

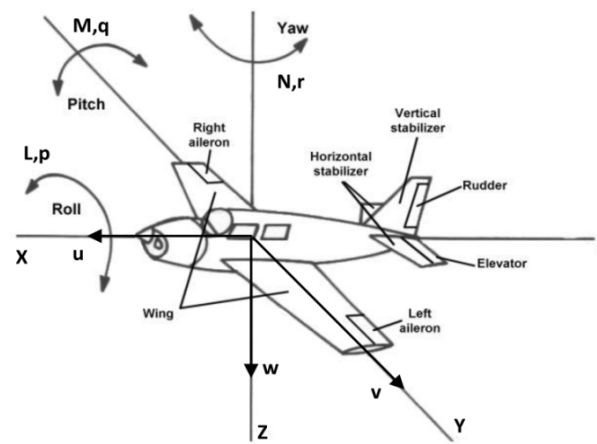

**Fig. 1** Direction of aircraft velocity vectors in relation to yaw, roll and pitch (this image is obtained from http://www.aerospaceweb.org)

## 2. Modeling

*1.1. Flight principles:*

Elevator, rudder and ailerons are three major actuators in guiding aircraft, which are being generally used. Yaw angle is controlled by the rudder on the vertical tail, and roll angle is controlled by ailerons on the wing tips. Pitch angle

**Table 2 Kinematic and dynamic equations**

| Kinematic and dynamic equations | |
|---|---|
| $X - mg \sin\theta = m(\dot{u} + qw - rv)$<br>$Y + mg \cos\theta \sin\Phi = m(\dot{v} + ru - pw)$<br>$Z + mg \cos\theta \cos\Phi = m(\dot{w} + pv - qu)$ | Force equations |
| $L = I_x \dot{p} - I_{xz}\dot{r} + qr(I_z - I_y) - I_{xz}pq$<br>$M = I_y \dot{q} + rq(I_x - I_z) + I_{xz}(p^2 - r^2)$<br>$N = -I_{xz}\dot{p} + I_z\dot{r} + pq(I_y - I_x) + I_{xz}qr$ | Moment equations |
| $p = \dot{\Phi} - \dot{\psi}\sin\theta$<br>$q = \dot{\theta}\cos\Phi + \dot{\psi}\cos\theta\sin\Phi$<br>$r = \dot{\psi}\cos\theta\cos\Phi - \dot{\theta}\sin\Phi$ | Body angular velocities in terms of Euler angles and Euler retes |
| $\dot{\theta} = q\cos\Phi - r\sin\Phi$<br>$\dot{\Phi} = p + q\sin\Phi\tan\theta + r\cos\Phi\tan\theta$<br>$\dot{\psi} = (q\sin\Phi + r\cos\Phi)\sec\theta$ | Euler retes in terms of Euler angles and body angular velocities |

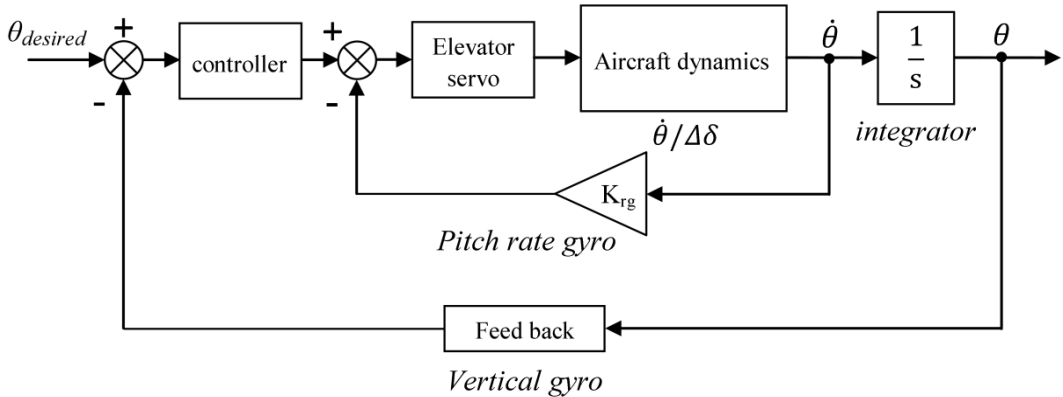

Fig. 2 the control system of pitch altitude is controlled by adjustment of elevator and is determined by aircraft turn around the transverse axis ($Y$). When elevator moves up, the aircraft will nose up and when it moves down, the aircraft will nose down. The direction of aircraft velocity vector in relation to yaw, roll and pitch is depicted in fig. 1.

*1.2. Mathematical model of pitch control*

A set of differential equations were used in order to describe the system's dynamics. These equations are obtained from development of Newton's second law, corresponding Figure1. Obtained force and moment equations are mentioned in table 2.

In mathematical modeling of system, motion equations must be linearized. First in modeling process, some assumptions are necessary to be considered: 1) the aircraft is assumed to be in steady state condition at constant altitude and velocity 2) changes of pitch angle do not alter the speed of aircraft. In this approach, small disturbance theory was used to linearize force and moment equations (Wahid & Rahmat, 2010). In the motion equations, all variables are replaced by a reference value plus a disturbance:

$u = u_0 + \Delta u \quad v = v_0 + \Delta v \quad w = w_0 + \Delta w$
$p = p_0 + \Delta p \quad q = q_0 + \Delta q \quad r = r_0 + \Delta r$
$X = X_0 + \Delta X \quad Y = Y_0 + \Delta Y \quad Z = Z_0 + \Delta Z$
$M = M_0 + \Delta M \quad N = N_0 + \Delta N \quad L = L_0 + \Delta L$
$\delta = \delta_0 + \Delta\delta$

For convenience, the reference flight condition is assumed to be symmetric and the propulsive forces are assumed to remain constant. Also if we initially align the x axis so that it is along the direction of velocity vector, then $w_0 = 0$ (Nelson, 1998). By these assumptions:

$v_0 = p_0 = q_0 = r_0 = \Phi_0 = \psi_0 = w_0 = 0$

Pitch belongs to the aircraft longitudinal motion. In this paper we only considered the pitch motion, thus longitudinal equations of motion were obtained through linearization:

$\left(\frac{d}{dt} - X_u\right)\Delta u - X_w \Delta w + (g\cos\theta_0)\Delta\theta = X_{\delta_e}\Delta\delta_e + X_{\delta_T}\Delta\delta_T$

$-Z_u \Delta u + \left((1 - Z_{\dot{w}})\frac{d}{dt} - Z_w\right)\Delta w -$

$\left((u_0 + Z_q)\frac{d}{dt} - g\sin\theta_0\right)\Delta\theta = Z_{\delta_e}\Delta\delta_e + Z_{\delta_T}\Delta\delta_T$

$-M_u \Delta u - \left(M_{\dot{w}}\frac{d}{dt} - M_w\right)\Delta w + \left(\frac{d^2}{dt^2} - M_q \frac{d}{dt}\right)\Delta\theta = M_{\delta_e}\Delta\delta_e + M_{\delta_T}\Delta\delta_T$

For obtaining the longitudinal transfer functions, Laplace transform will be used. All initial conditions set to zero (Nelson, 1998). First we assume:

$\theta_0 = 0 \rightarrow \cos\theta_0 = 1 \quad \sin\theta_0 = 0 \quad Z_q = Z_{\dot{w}}$

The following set of longitudinal differential equations is yielded by incorporating all of aforementioned assumptions:

$\left(\frac{d}{dt} - X_u\right)\Delta u - X_w \Delta w + g\Delta\theta = X_{\delta_e}\Delta\delta_e$
$-Z_u \Delta u + \left(\frac{d}{dt} - Z_w\right)\Delta w - u_0 \frac{d}{dt}\Delta\theta = Z_{\delta_e}\Delta\delta_e$
$-M_u \Delta u - \left(M_{\dot{w}}\frac{d}{dt} - M_w\right)\Delta w + \frac{d}{dt}\left(\frac{d}{dt} - M_q\right)\Delta\theta = M_{\delta_e}\Delta\delta_e$

After taking Laplace transformation, the transfer function of the pitch angle to elevator deflection is as follow:

$\frac{\overline{\Delta\theta}}{\overline{\Delta\delta_e}} = \frac{A_\theta s^2 + B_\theta s + C_\theta}{As^4 + Bs^3 + Cs^2 + Ds + E}$

The fig. 2 depicts the pitch altitude control system. The reference pitch angle is compared with actual angle that is measured by *vertical gyro* (external loop) to produce an error signal for activating the control servo. In traditional schemes this error signal is then amplified (by proportional gain) and sent to the control surface actuator (*elevator servo*) to deflect it. The inner feedback loop is adopted for damping the oscillations. Movement of elevator actuator makes the aircraft to obtain a new pitch orientation (Nelson, 1998). The elevator servo transfer function can be displayed as follows:

$$\frac{\delta_e}{v} = \frac{1}{\tau s + 1}$$

Where $\delta_e, v,$ and $\tau$ are the elevator deflection angle, input voltage, and servomotor time constant, respectively (Nelson, 1998).

The Boeing 747-400 parameters were used to achieve transfer functions of the *pitch angle to elevator deflection* and *elevator servo* (Barros dos Santos & de Oliveira, 2011):

$\frac{\overline{\Delta\theta}}{\overline{\Delta\delta_e}} = \frac{-1.69144s^2 - 0.84341s - 0.0099096}{s^4 + 1.17103s^3 + 1.55405s^2 + 0.012538s + 0.0072771}$

$\frac{\delta_e}{v} = \frac{10}{s + 10}$

### 3. Design:

#### 3.1. Structure and function of FSPID controller:

FSPID includes two parts: 1) adjustable PID controller, and 2) fuzzy inference mechanism, as displayed in fig .3.

Error "$e$" and changes-in-error "$ec$" were obtained as input of fuzzy inference mechanism. Then fuzzy inference mechanism uses adjustment law (1) to explore the fuzzy relationship between PID parameters with "$e$" and "$ec$".

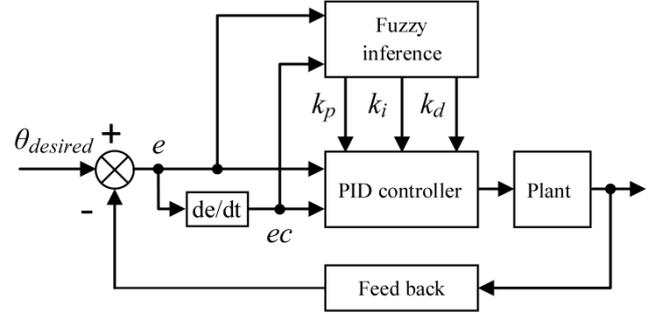

**Fig. 3 Fuzzy self-tuning PID controller**

$k_p = k'_p + \Delta k_p$
$k_i = k'_i + \Delta k_d \quad (1)$
$k_d = k'_d + \Delta k_d$

$\Delta k_p$, $\Delta k_i$, and $\Delta k_d$ are outputs of fuzzy controller in the above law. Initial values of PID controller are $k_p'$, $k_i'$, and $k_d'$. Then $k_p$, $k_i$, and $k_d$, will be tuned to provide online and adaptive self-tuning in different conditions of system function.

#### 3.2. Automatic parameter-tuning rules of PID controller:

The conventional PID controller equation is:

$U(K) = k_p E(K) + k_i \Sigma E(k) + k_d EC(K)$

The error of input variable and the changes-in-error in the equation are respectively $\Sigma E(k) = E(k) + E(k-l)$ and $EC(k) = E(k) - E(k-l)\ (k = 0,1,2)$. Parameters that characterize the proportion, integral and differential role are respectively $k_p$, $k_i$, and $k_d$.

The proportional coefficient $k_p$ improves the response speed of system and adjustment precision. A note to be considered is that in excessive amounts of $k_p$, overshoot and even system instability will be made (Zhao & Pan, 2010). The integral coefficient $k_i$ diminishes steady-state error of system. However, immoderate $k_i$ can lead to integral saturation and overshoot. The function of derivative coefficient $k_d$ is to improve dynamic characteristics of system. Thus bigger $k_d$ amounts prevent changes of error in different directions over the response process. Again, very large $k_d$ causes prolonged adjustment time and reduced anti-interference performance (Zhao & Pan, 2010).

These are fundamental rules for automatic tuning of PID parameters in accord to the impact of the parameters $k_p$, $k_i$, and $k_d$ considering different errors and changes-in-error:

1) When "$e$" is large, greater $k_p$ should be chosen to improve the system response speed, also $k_d$ should be taken small for avoiding the differential over saturation. In rapidly increase of "$e$", $k_i$ should be very small even zero to prevent integral saturation and big overshoot (Yongbin, et al., 2010).

2) In the middle amounts of "$e$" and "$ec$", appropriate $k_i$ and $k_d$ in addition to a smaller amount of $k_p$ should be chosen to reduce overshoot and improve system's response speed (Yongbin, et al., 2010).

**Table 3 Fuzzy Rules**

| ec | NB | NM | NS | ZO | PS | PM | PB |
|---|---|---|---|---|---|---|---|
| e | $k_p/k_i/k_d$ | | | | | | |
| NB | PB/NB/PS | PB/NB/NM | PM/NB/NB | PM/NM/NB | PS/NS/NB | PS/ZO/NM | ZO/ZO//PS |
| NM | PB/NB/PS | PB/NB/NS | PM/NM/NB | PS/NS/NM | PS/NS/NM | ZO/ZO/NS | NS/ZO/ PS |
| NS | PM/NB/ZO | PM/NM/NS | PM/NS/NM | PS/NS/NM | ZO/ZO/NS | NS/PS/NS | NM/PS/ZO |
| ZO | PM/NM/ZO | PM/NS/NS | PS/NS/NS | ZO/ZO/NS | NS/PS/NS | NM/PM/NS | NM/PM/ZO |
| PS | PS/NS/ZO | PS/NS/NS | ZO/ZO/ZO | NS/PS/ZO | NS/PS/ZO | NM/PM/ZO | NM/PB/PS |
| PM | ZO/ZO/PB | ZO/ZO/NS | NS/PS/PS | NM/PS/PS | NM/PM/PS | NM/PB/PS | NB/PB/PB |
| PB | ZO/ZO//PB | NS/ZO//PM | NM/PS/PM | NM/PM/PM | NM/PB/PS | NB/PB/PS | NB/PB/PB |

3) When "$e$" is small, greater $k_p$ and $k_i$ should be chosen to have a better steady state performance. At the same time, in light of disturbance-resisting ability of system the proper $k_d$ should be appointed for avoiding oscillations of system. When "$ec$" is big, $k_d$ should be smaller and when "$ec$" is small, $k_d$ should be bigger (Yongbin, et al., 2010).

### 3.3. Determine membership functions and fuzzy rules

#### 3.3.1. Fuzzification of input and output variables

Based on the fuzzy set theory, first the input and output values should be transformed into linguistic variables, which is called *fuzzification*. The ranges of input and output variables are $e, ec \in [-5, 5], k_p, k_i, k_d \in [-5, 5]$. Then the fuzzy range of input and output values was divided into 7 linguistic variables. These fuzzy subsets are:

$e, ec = NB, NM, NS, ZO, PS, PM, PB$
$k_p, k_i, k_d = NB, NM, NS, ZO, PS, PM, PB$

Where $NB$ is negative big; $NM$ is negative medium; $NS$ is negative small; $ZO$ is zero; $PS$ is positive small; $PM$ is positive medium; and $PB$ is positive big.

Gaussian and triangular membership functions were used in inputs and outputs, respectively. Fig. 4. represents the membership functions.

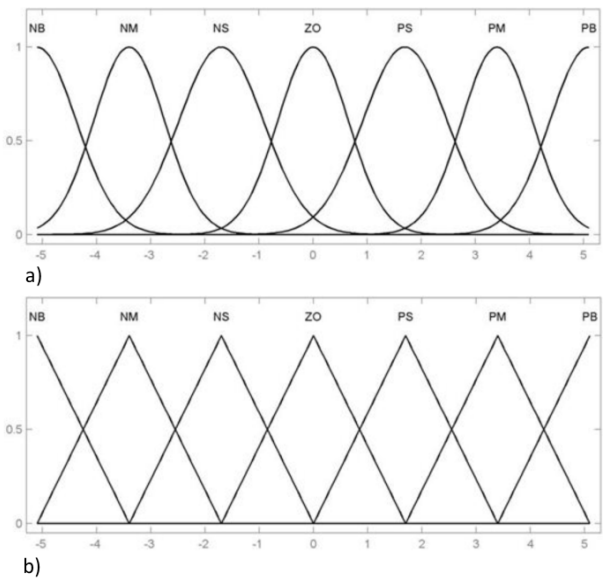

**Fig. 4 Membership Functions, a) inputs, b) outputs**

#### 3.3.2. Establishing fuzzy rules

Table. 3 shows the fuzzy rules based on automatic parameter-tuning rules of PID controller that was explained before. These rules are obtained by trial-and-error method in addition to expert knowledge. The fuzzy reasoning rules are expressed in this form:

*If e is $A_i$ and ec is $B_j$; then $k_p$ is $C_{ij}$; $k_i$ is $D_{ij}$ and $k_d$ is $E_{ij}$*

Where $A_i, B_j, C_{ij}, D_{ij}, E_{ij}$ are fuzzy subsets of inputs and outputs, and

$i, j = 1,2,3,4,5,6,7$.

#### 3.3.3. Fuzzy inference and defuzzification

In this study Mamdani's inference method was manipulated. In this way, min and max operators were gotten in order to accomplish fuzzy outputs:

$$\mu_{c'}(k_p) = \bigvee_{i,j=1}^{7} \{[\mu_i(e) \wedge \mu_j(ec)] \wedge \mu_{c_{ij}}(k_p)\}$$

$k_i$ and $k_d$ were similarly caught.

The centroid method (center of gravity) was used for defuzzificating of outputs:

$$k_p(e,ec) = \frac{\sum_{n=1}^{N} k_{pn} \mu_{c'}(k_{pn})}{\sum_{n=1}^{N} \mu_{c'}(k_{pn})}$$

### 3.4. Design of autopilot controllers

Amplifier is a proportional controller (PC) in conventional autopilot applications (Nelson, 1998) and is displayed in the fig 2. To compare control features, PC was replaced by both CPID and proposed controller FSPID. Finally the outcomes of these three methods will be discussed.

Inner feedback loop (in fig. 2) functions as oscillation damper. Then the inner feedback loop (pitch rate gyro) was removed in order to compare the function of CPID, FSPID and PC in the un-damped oscillations. In this usage, two different tunings were used for CPID.

Finally, the short-period approximation was applied and the efficacy of both CPID and FSPID in face with disturbance is examined. We intentionally manipulated inappropriate tunings for PID to explore the influence of FSPID on the response speed of system and overshoots. The short-period transfer function (Wahid & Rahmat, 2010) is:

$$\frac{\overline{\Delta\theta}}{\overline{\Delta\delta_e}} = \frac{11.7304s+22.578}{s^3+4.9676s^2+12.941s}$$

In simulations of short-period approximations, the ranges of input and output variables are $e, ec \in [-40,40], k_p, k_i, k_d \in [-40,40]$.

### 4. Simulation:

Fig. 5 describes the difference between control features of three controllers which mentioned above. Firstly discouraging outcomes of PC can be clearly observed as the big overshoot (14.6%) compared to FSPID (0%), and visible steady-state error (2%) compared to FSPID (0%). Also CPID has 3% steady-state error. Conventional setting of CPID (fixed-gain) does not conduct the optimal response. Results of Fuzzy inference demonstrate a better tuning for PID controller as is shown in fig. 5.

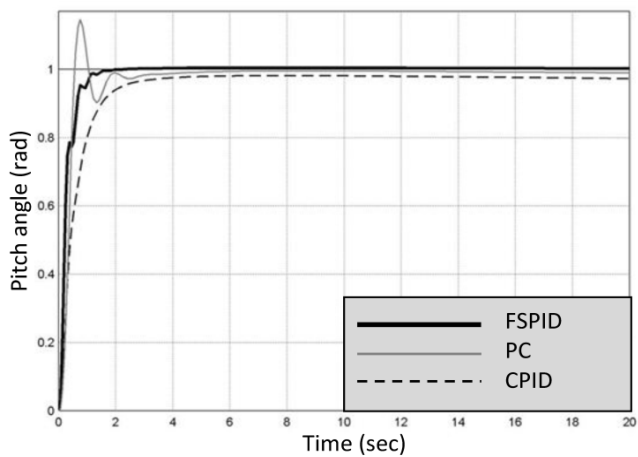

Fig. 5 Comparison of the controllers (PC, CPID, and FSPID).

Removing the inner feedback loop causes un-damped oscillation for PC as is shown in fig. 6. Both the FSPID controller and CPID show a better function in lack of inner feedback loop (*gyro*) in comparison with PC. While both of CPID tunings has big overshoots (22.8% and 31%), FSPID has not.

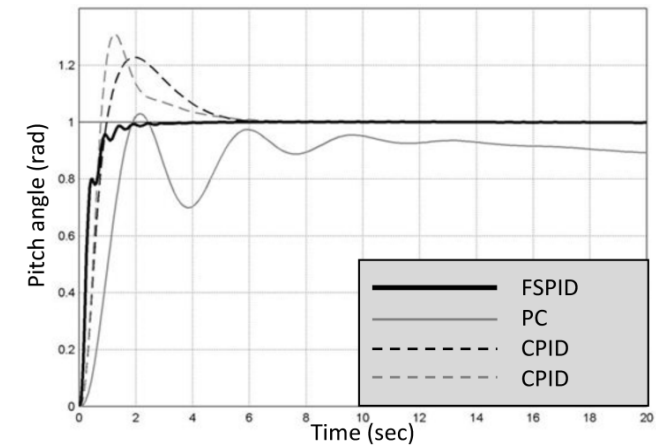

Fig. 6 Different functions between FSPID, CPID, and PC in lack of inner feedback loop.

Simulation results of these controllers in short-period approximation are illustrated in fig. 7, 8, and 9. Advantages of FSPID can be clearly implied in comparison with CPID.

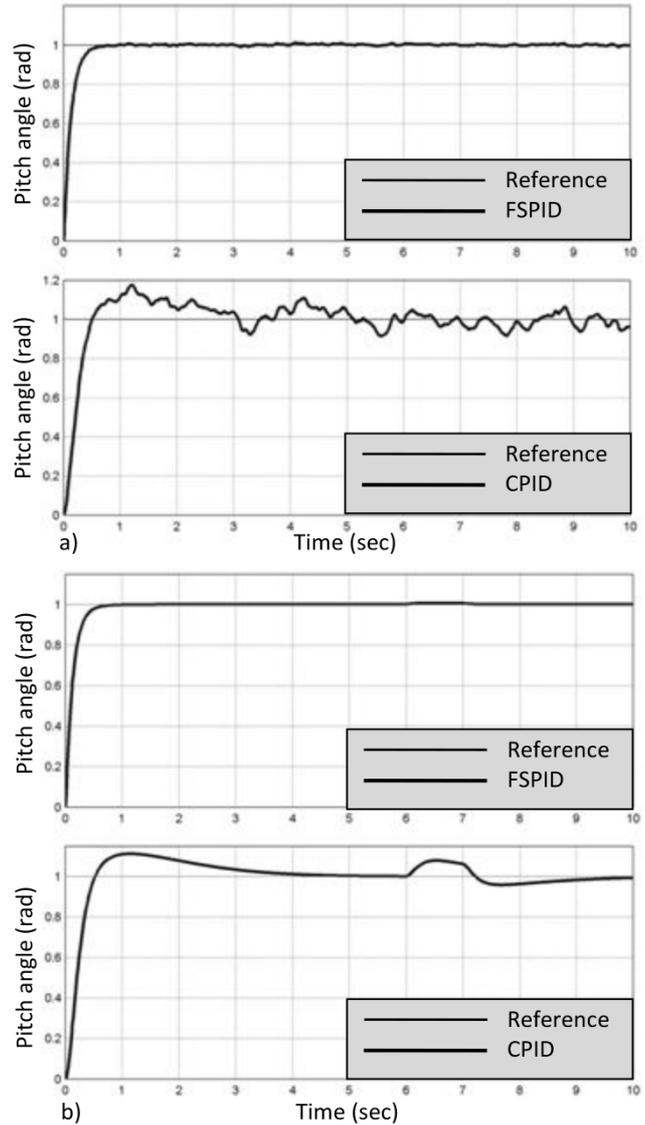

Fig. 7 a) continuous disturbance, b) abrupt disturbance

Abrupt and continuous disturbances were manipulated (they can assumed as unfavorable atmosphere conditions which are common in aviation). Fig. 7 compares the anti-disturbance function of FSPID and CPID. The simulation results demonstrate that the anti-disturbance function of FSPID was more successful compared to conventional PID.

Refer to fig. 8a, In FSPID rise time reduces from 2.3 s to 1.09 s, a gradually improvement (53%) in response speed. However, In CPID rise time increases from 2.32 s to 2.5 s (about 7% deterioration in speed). In accord to these simulations CPID is not adaptive because of using fixed-gain, so that it cannot improve the system response for example in conditions of inappropriate setting of PID parameters or long-term alterations in system parameters. The fig. 8b compares the ability of CPID and FSPID to decrease the overshoot. FSPID could decrease the overshoot (about 5.9%) to zero and concurrently accelerated (48%) the system response speed (rise time reduces from 1.4 s to 0.73 s). CPID reduces overshoot from 6.2% to 3.3% so cannot completely eliminate it, concurrently rise time is increased from 1.4 s to 1.72 (18% deterioration in speed).

Fig. 9 describes the tracking ability of CPID and FSPID. Both of these controllers can desirably track the commands, but only FSPID can optimize the tracking function so that the rise time reduces from 1.38 s to 0.84 s (40% improvement in speed) and in the same time overshoot reduces from 1.76% to zero. However, In CPID rise time increases from 1.5 s to 1.73 s (13.3% deterioration in speed) and a constant 2.2% overshoot can be seen.

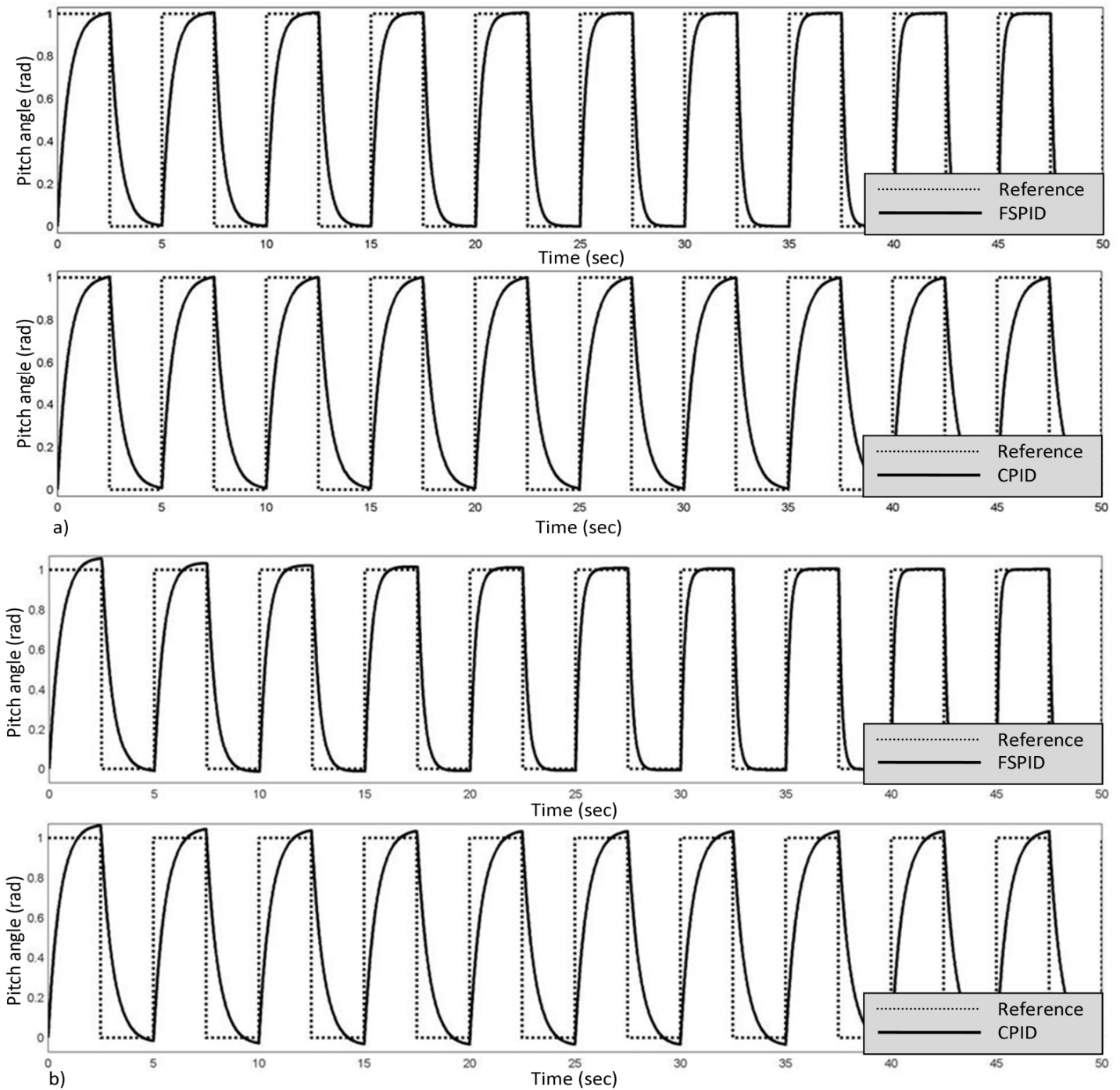

**Fig. 8** Comparisons between FSPID and CPID in conditions of inappropriate setting of PID parameters: a) speed of system response; b) Overshoot plus speed of system response

**5. Conclusion:**

We conclude that FSPID had adaptive features so that it could tune the PID parameters in an online process for achieving the best response in terms of speed, overshoot, and steady-state error. The function of CPID was better than PC, but was optimized by our proposed controller. Although traditional schemes of autopilots need inner feedback loop to damp the oscillations, the CPID and FSPID could properly damp these oscillations in lack of inner feedback loop. FSPID could optimally overcome the abrupt and continuous disturbances, based on the simulation comparisons between CPID and FSPID in the autopilot controller. The function of FSPID in face with inappropriate setting of PID parameters or conditions such as long-term changes of system parameters was perfect so that it could improve the system response in view of speed and overshoot. CPID did not show this ability because of using fixed-gain.

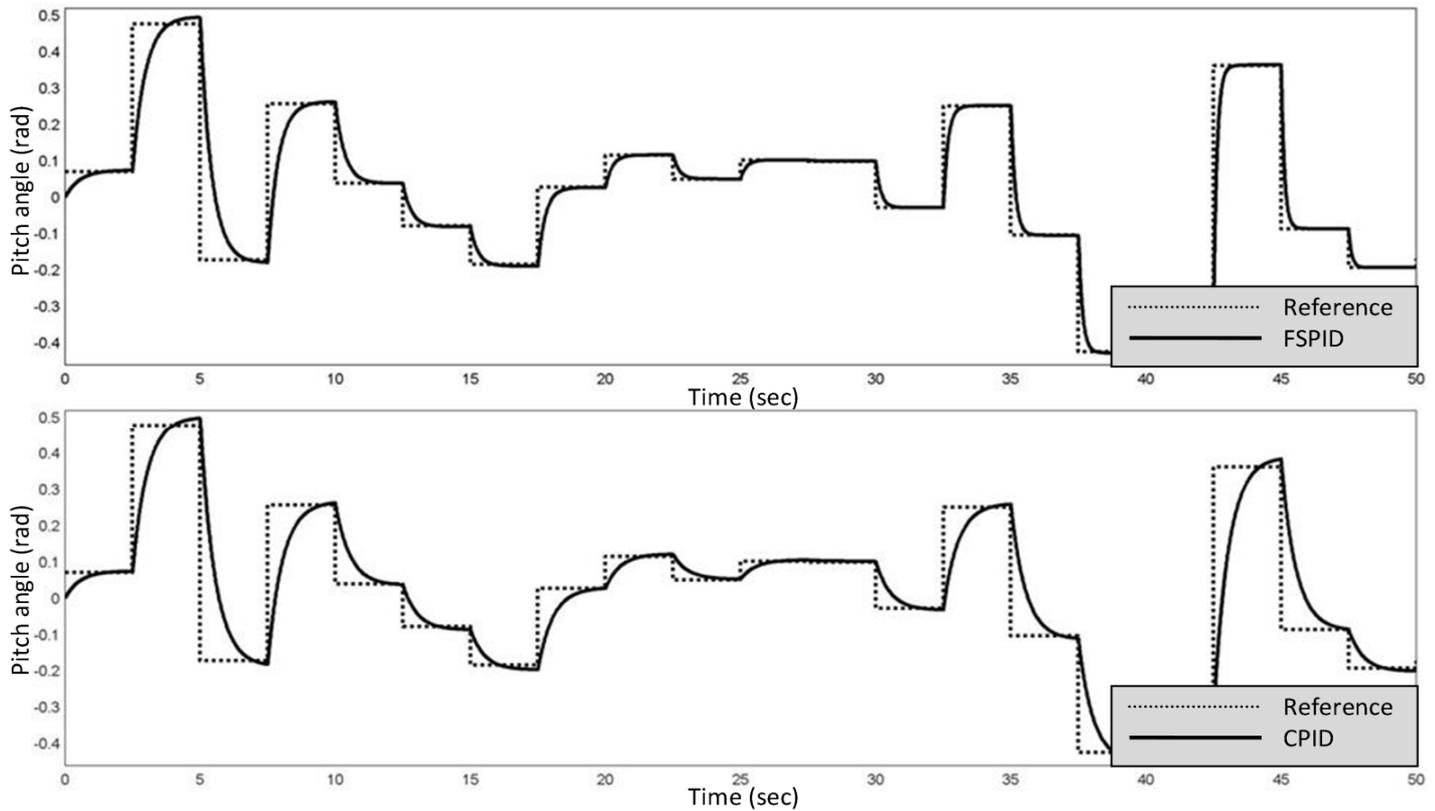

**Fig. 9 Tracking of commands: FSPID eliminates overshoot and improves speed response.**

Our fuzzy inference was designed based on trial-and-error plus expert knowledge. This proposed method in autopilot control can be improved in future studies by employing intelligent methods such as genetic algorithm or neural network.

**Acknowledgement:**

We are indepted to Mr Hamid Reza Goharian, member of power electronic lab of Tarbiat Modares University, for helpful suggestions and supports